\documentclass[twocolumn,
               amsmath,
               amssymb,
               aps,
               superscriptaddress,
               longbibliography,
               10pt]
               {revtex4-2}

\usepackage{graphicx}
\usepackage{dcolumn}
\usepackage{bm}
\usepackage{color}
\usepackage{url}
\usepackage{tabularx}
\usepackage{braket}
\usepackage{comment}
\usepackage{here}
\usepackage{algorithm}
\usepackage{algpseudocode}
\usepackage{booktabs}
\usepackage{siunitx}
\usepackage{subfiles}
\usepackage[colorlinks=true,citecolor=blue,linkcolor=blue,urlcolor=blue]{hyperref}

\begin{document}
\title{Fair sampling of ground-state configurations using hybrid quantum-classical MCMC algorithms}

\author{Yuichiro Nakano}
\email{u830977g@ecs.osaka-u.ac.jp}
\affiliation{
    Graduate School of Engineering Science, The University of Osaka, 1-3 Machikaneyama, Toyonaka, Osaka 560-8531, Japan.
}

\author{Keisuke Fujii}
\email{fujii.keisuke.es@osaka-u.ac.jp}
\affiliation{
    Graduate School of Engineering Science, The University of Osaka, 1-3 Machikaneyama, Toyonaka, Osaka 560-8531, Japan.
}
\affiliation{
    Center for Quantum Information and Quantum Biology, The University of Osaka, 1-2 Machikaneyama, Toyonaka, 560-0043, Japan.
}
\affiliation{
    Center for Quantum Computing, RIKEN, Wako Saitama 351-0198, Japan.
}

\date{\today}

\begin{abstract}
We study the fair-sampling properties of hybrid quantum-classical Markov chain Monte Carlo (MCMC) algorithms for combinatorial optimization problems with degenerate ground states. 
While quantum optimization heuristics such as quantum annealing and the quantum approximate optimization algorithm (QAOA) are known to induce biased sampling, hybrid quantum-classical MCMC incorporates quantum dynamics only as a proposal transition and enforces detailed balance through classical acceptance steps.
Using small Ising models, we show that MCMC post-processing corrects the sampling bias of quantum dynamics and restores near-uniform sampling over degenerate ground states. 
We then apply the method to random $k$-SAT problems near the satisfiability threshold. For random 2-SAT, a hybrid MCMC combining QAOA-assisted neural proposals with single spin-flip updates achieves fairness comparable to that of PT-ICM. 
For random 3-SAT, where such classical methods are no longer applicable, the hybrid MCMC still attains approximately uniform sampling. We also examine solution counting and find that the required number of transitions is comparable to that of WalkSAT.
These results indicate that hybrid quantum-classical MCMC provides a viable framework for fair sampling and solution enumeration.
\end{abstract}

\maketitle


\label{introduction}
\section{introduction}
Combinatorial optimization problems play an indispensable role across a wide range of fields, including logistics, finance, drug discovery, and machine learning \cite{Korte2018}. However, many practically important problems are classified as NP-hard in computational complexity theory, meaning that their computational cost typically grows exponentially with problem size, which makes the computation of exact optimal solutions extremely challenging. Consequently, a great deal of effort has long been devoted to the development of approximation algorithms and heuristics that can provide high-quality approximate solutions within practical time scales \cite{Williamson_Shmoys_2011}. In recent years, rapid advances in quantum hardware have spurred interest in new optimization heuristics that exploit the principles of quantum mechanics. Among these, quantum annealing (QA) \cite{kadowaki1998quantum, farhi2000quantumcomputationadiabaticevolution} and the quantum approximate optimization algorithm (QAOA) \cite{farhi2014quantum} are regarded as two of the most widely studied and promising approaches. In both methods, the cost function of an optimization problem is encoded into a classical Ising Hamiltonian $H_P$, which is combined with a driver Hamiltonian $H_d$ (such as a transverse field) that induces quantum superposition and transitions between configurations, thereby enabling exploration of the solution space. QA aims to reach the ground state of $H_P$ (the optimal solution) through adiabatic time evolution implemented on a quantum annealer, whereas QAOA pursues the same goal using parameterized quantum circuits on universal quantum computers. These quantum heuristics have been applied to a variety of optimization tasks, and numerical evidence has suggested potential advantages in time-to-solution scaling over classical algorithms under specific conditions \cite{PRXQuantum.5.030348, PhysRevLett.134.160601}, positioning them as promising candidates for next-generation computational paradigms.

On the other hand, in several important applications of optimization, it is not sufficient to find a single optimal solution. Instead, one must exhaustively explore the entire set of optimal solutions (ground states). Representative examples include model counting problems (\#SAT) \cite{VALIANT1979189}, counting solutions of knapsack instances (\#Knapsack) \cite{6108252}, and SAT-based membership filters \cite{weaver2012satisfiability, fang2018naesatbased}. For such tasks, the ability to draw solutions uniformly at random from the solution space, that is, \emph{fair sampling}, is of critical importance. However, the standard transverse-field driver widely used in QA and QAOA, $H_d = -\sum_i \sigma_i^x$, is known to induce biased transition probabilities due to quantum interference effects, leading to an exponential suppression of the occurrence probabilities of certain ground states \cite{Matsuda_2009, PhysRevLett.118.070502, PhysRevA.100.030303, PhysRevE.111.054103}. To mitigate this bias and achieve perfectly fair sampling, the use of higher-order interaction drivers has been proposed in QA \cite{Matsuda_2009}, while in QAOA, the introduction of Grover-mixers has been suggested \cite{sundar2019quantumalgorithmcountweighted, 9259965, Zhu_2023, Zhang_2025}. Nonetheless, implementing these more complex drivers typically entails substantial overhead in terms of hardware connectivity and circuit depth, rendering them impractical on current noisy intermediate-scale quantum (NISQ) devices \cite{PhysRevA.100.030303}.

Against this backdrop, there has been growing interest in leveraging quantum optimization heuristics not only as solvers for optimization problems but also as building blocks for sampling from probability distributions. A particularly promising line of research in this direction is hybrid quantum-classical Markov chain Monte Carlo (MCMC) \cite{layden2023quantum}. In this approach, samples drawn from a quantum circuit are used as a proposal distribution in the transition steps of MCMC methods, such as the Metropolis--Hastings algorithm. Previous work has demonstrated that transitions driven by a time-independent QA Hamiltonian can be effective for sampling from low-temperature Boltzmann distributions of Ising models \cite{layden2023quantum}. In addition, methods have been proposed that employ autoregressive neural-network samplers trained to approximate the output distribution of QAOA circuits, enabling adaptive sampling \cite{10.21468/SciPostPhys.15.1.018, nakano2025neuralnetworkassistedmontecarlosampling}. These hybrid schemes aim to enhance sampling efficiency by combining the local exploration capabilities of classical MCMC with the global exploration abilities provided by QA and QAOA. This naturally leads to a central open question: \emph{Can hybrid quantum-classical MCMC with a standard transverse-field driver achieve fair sampling while retaining the practical advantages of easy hardware implementation?} If the sampling bias can indeed be mitigated within the MCMC framework using such a simple driver, without resorting to more elaborate mixers, it would constitute a highly practical and significant sampling algorithm for near-term quantum devices.

In this work, we investigate the fair-sampling capability of hybrid quantum-classical MCMC. 
Specifically, we explore its fair-sampling performance and applications through two problem settings based on Ising models. 
The first is an analysis of small-scale Ising models, including toy models studied in previous work. 
There we show that introducing MCMC steps on top of the quantum dynamics can restore fair sampling. 
Furthermore, as a more practically relevant problem setting, we consider sampling solutions of random $k$-SAT instances. 
For $k = 2$, we find that combining proposals generated by quantum dynamics with a standard single spin-flip (SSF) update achieves a level of fairness comparable to that of PT-ICM \cite{zhu2015efficient}, which is known to realize fair sampling. 
For $k = 3$, where sophisticated classical sampling algorithms such as PT-ICM are no longer applicable, the hybrid quantum–classical MCMC still attains fair sampling. 
In addition, for solution counting, we show that it requires a number of transitions comparable to that of WalkSAT \cite{10.5555/1597148.1597256}, a representative classical heuristic solver. 
These results point to new possibilities for applying quantum heuristics to solution-sampling tasks, which have traditionally been regarded as a weak point of such methods.

The remainder of this paper is organized as follows. 
In Sec.~\ref{sec:model_and_method}, we introduce the algorithms compared in this study. 
Section~\ref{sec:fair_sampling} examines the fair-sampling capability using representative small-scale instances that have been employed in previous work. 
Section~\ref{sec:random_sat_sampling} investigates the sampling of solutions for random $k$-SAT problems, comparing our approach with classical algorithms and discussing applications to counting problems. 
Finally, we present our conclusions and discuss future prospects in Sec.~\ref{sec:conclusion}.


\section{Model \& Method}
\label{sec:model_and_method}
\subsection{Ising model}
The Ising model is a mathematical framework originally proposed in statistical physics to describe phase transitions in ferromagnetic materials \cite{Ising1925, RevModPhys.39.883}. In this model, the magnetic spin of each atom is represented by a binary variable $s_i \in \{ +1, -1 \} $, which corresponds to the up ($+1$) or down ($-1$) spin state. 
In general, the energy of a model with at most $k$-body spin interactions, or Hamiltonian, is given by
\begin{align}
H_P(\boldsymbol{s}) = \sum_{q=1}^{k} \sum_{1 \le i_1 < i_2 < \dots < i_q \le N} J_{i_1 i_2 \dots i_q}^{(q)} s_{i_1} s_{i_2} \dots s_{i_q},   
\end{align}
where $J_{i_1 \dots i_q}^{(q)}$ denotes the interaction coefficient among the $q$-spins $(s_{i_1}, \dots, s_{i_q})$.
In combinatorial optimization, the Ising model is intimately linked to problems defined by objective functions that are polynomials of degree $k$ over discrete variables.
By mapping the objective function onto the energy function $H_P(\boldsymbol{s})$, the problem of minimizing the objective function reduces to identifying the physical state with the lowest energy, known as the ground state.
The special case ($k = 2$) corresponds to the quadratic unconstrained binary optimization (QUBO) formulation, a widely studied standard model in combinatorial optimization \cite{kochenberger2014, 10.3389/fphy.2014.00005}. When all couplings $J_{ij}$ are negative, corresponding to ferromagnetic interactions, the trivial configuration in which all spins align in the same direction minimizes the energy. However, in realistic optimization problems, the signs of the couplings are typically mixed. This leads to frustration, where competing local energy minimization conditions cannot be satisfied simultaneously. As a result, the model exhibits a complex energy landscape with many local minima \cite{PhysRevLett.35.1792, doi:10.1142/1655, RevModPhys.58.801}. Finding the ground state of such a general Ising model is classified as NP-hard \cite{Barahona_1982}, and developing efficient search algorithms remains a significant challenge.

On the other hand, in statistical physics, the probability that a system in thermal equilibrium at temperature $T$ occupies a configuration $\boldsymbol{\sigma}$ is given by the Boltzmann distribution,
\begin{align}
\pi(\boldsymbol{\sigma}) = \frac{e^{-\beta H(\boldsymbol{\sigma})}}{Z(\beta)} ,
\end{align}
where $\beta = 1/(k_B T)$ is the inverse temperature with $k_B$ being the Boltzmann constant, and $Z(\beta) = \sum_{\boldsymbol{\sigma}} e^{-\beta H(\boldsymbol{\sigma})}$ is the partition function. This distribution implies that configurations with lower energy $H(\boldsymbol{\sigma})$ appear with exponentially higher probability. In the zero-temperature limit ($T \to 0$, equivalently $\beta \to \infty$), the distribution becomes concentrated on the ground-state configurations $ \boldsymbol{\sigma}_{\mathrm{GS}}$ with minimum energy:
\begin{align}
\lim_{\beta \to \infty} \pi(\boldsymbol{\sigma}) =
\begin{cases}
1/N_g & \text{if } H(\boldsymbol{\sigma}) = E_{\min}, \\
0   & \text{otherwise},
\end{cases}
\end{align}
where $N_g$ is the ground-state degeneracy. 

Therefore, there are primarily two approaches to searching for the ground state of the Ising model:
(i) Optimization based on energy minimization. A traditional method in this category is Simulated Annealing (SA) \cite{doi:10.1126/science.220.4598.671}. This approach utilizes thermal fluctuations and gradually decreases the temperature, aiming to eventually reach the ground state.
(ii) Boltzmann distribution sampling at zero (or extremely low) temperatures. While similar to SA in that the temperature is gradually lowered, this approach aims to discover the ground state by directly sampling from the low-temperature Boltzmann distribution using sampling algorithms \cite{PhysRevE.92.013303}.
Indeed, one of the best classical optimization heuristics currently available for QUBO problems is known to be PT-ICM \cite{zhu2015efficient}, which is an efficient sampling algorithm for low-temperature Boltzmann distributions \cite{10.3389/fphy.2019.00048, PhysRevE.99.063314}.

\subsection{QA and QAOA}
In this section, we describe quantum annealing (QA) \cite{kadowaki1998quantum} and the quantum approximate optimization algorithm (QAOA) \cite{farhi2014quantum}, which are representative quantum heuristics for optimization problems that can be expressed using the Ising model.

We first represent the combinatorial optimization problem in the form of an Ising model. The objective function is mapped to the Hamiltonian $H_P$ of a quantum system, referred to as the problem Hamiltonian, which is constructed so that its ground state corresponds to the optimal solution. In general, $H_P$ can be written using Pauli-Z operators $\hat{\sigma}^z$ as
\begin{align}
H_P = \sum_{q=1}^{k} \sum_{1 \le i_1 < i_2 < \dots < i_q \le N} J_{i_1 i_2 \dots i_q}^{(q)} \hat{\sigma}_{i_1}^z \hat{\sigma}_{i_2}^z \dots \hat{\sigma}_{i_q}^z, 
\end{align}

Quantum annealing (QA) \cite{kadowaki1998quantum} is a method that uses quantum fluctuations to search for the ground state of $H_P$. One prepares an initial Hamiltonian $H_{d}$ with a trivial ground state, typically a transverse-field term $H_{d} = - \sum_i \hat{\sigma}_i^x$, and gradually interpolates the total Hamiltonian $H(t)$ from $H_{d}$ to $H_P$ as a function of time $t$:
\begin{align}
H(t) = A(t) H_{d} + B(t) H_P.
\end{align}
According to the adiabatic theorem, if this evolution is sufficiently slow, a system that begins in the ground state of $H_{d}$ remains in the instantaneous ground state of $H(t)$ throughout the process. As a result, at the end of the annealing schedule the system reaches the ground state of $H_P$, which yields the optimal solution.

The quantum approximate optimization algorithm (QAOA) \cite{farhi2014quantum} is a variational quantum algorithm for solving combinatorial optimization problems on universal quantum computers. It can be interpreted as a time-discretized counterpart of the adiabatic evolution used in QA. In QAOA, the time-evolution operator in QA is approximated, in a manner similar to the Suzuki–Trotter decomposition, by an alternating sequence of unitaries corresponding to evolution under $H_P$ (phase separation) and under $H_{d}$ (mixing). Specifically, in a depth-$p$ quantum circuit, the variational state $|\psi(\boldsymbol{\gamma}, \boldsymbol{\beta})\rangle$ is prepared as
\begin{align}
|\psi(\boldsymbol{\gamma}, \boldsymbol{\beta})\rangle
= \prod_{k=1}^{p} e^{-i \beta_k H_{d}} e^{-i \gamma_k H_P} |+\rangle^{\otimes N},    
\end{align}
where $|+\rangle^{\otimes N}$ is the ground state of $H_{d}$, that is, the uniform superposition state, and $\boldsymbol{\gamma} = (\gamma_1, \dots, \gamma_p)$ and $\boldsymbol{\beta} = (\beta_1, \dots, \beta_p)$ are variational parameters. Unlike QA, which varies the Hamiltonian continuously in real time, QAOA optimizes the parameters $ (\boldsymbol{\gamma}, \boldsymbol{\beta})$ using classical optimization methods to minimize the expectation value $\langle \psi | H_P | \psi \rangle$, thereby searching for an approximate solution. In the limit $p \to \infty$, QAOA is known to have sufficient expressive power to reproduce adiabatic quantum computation \cite{farhi2000quantumcomputationadiabaticevolution}.

\subsection{Hybrid quantum-classical MCMC}
In this section, we introduce the hybrid quantum-classical MCMC algorithm, which serves as a quantum heuristic sampler for Ising models.

To evaluate the Boltzmann distribution $\pi(\boldsymbol{\sigma})$ directly, one must compute the partition function $Z(\beta) = \sum_{\boldsymbol{\sigma}} e^{-\beta H(\boldsymbol{\sigma})}$.
However, as the number of spins $N$ increases, the size of the state space grows exponentially as $2^N$, making the exact computation of $Z(\beta)$, which requires summation over all configurations, practically impossible. Markov chain Monte Carlo (MCMC) methods \cite{metropolis1953equation, hastings1970monte} address this difficulty by approximately generating a sequence of samples $\{\boldsymbol{\sigma}^{(1)}, \boldsymbol{\sigma}^{(2)}, \dots\}$ drawn from the Boltzmann distribution.

In MCMC, a Markov chain is constructed in which the system transitions stochastically from the current state $\boldsymbol{\sigma}$ to a new state $\boldsymbol{\sigma}'$. If the transition probability $P(\boldsymbol{\sigma}' | \boldsymbol{\sigma})$ satisfies the detailed balance condition
\begin{align}
\pi(\boldsymbol{\sigma}) P(\boldsymbol{\sigma}'| \boldsymbol{\sigma})
= \pi(\boldsymbol{\sigma}') P(\boldsymbol{\sigma}| \boldsymbol{\sigma}')
\quad \forall \boldsymbol{\sigma}, \boldsymbol{\sigma}',
\end{align}
then the stationary distribution of the Markov chain is guaranteed to converge to the target distribution $\pi(\boldsymbol{\sigma})$. This condition expresses reversibility in thermal equilibrium, and by designing the transition rule to satisfy it, one can sample from the desired distribution without explicitly computing $Z(\beta)$.

The Metropolis--Hastings algorithm \cite{hastings1970monte} is a general framework for constructing Markov chains that satisfy detailed balance. Its central idea is to decompose each transition probability into a proposal distribution and an acceptance probability:
\begin{align}
P(\boldsymbol{\sigma}' | \boldsymbol{\sigma}) =
\begin{cases}
Q(\boldsymbol{\sigma}'|\boldsymbol{\sigma})A(\boldsymbol{\sigma}'|\boldsymbol{\sigma}) & \text{if } \boldsymbol{\sigma}' \neq \boldsymbol{\sigma}, \\
1 - \sum_{\boldsymbol{\sigma}''\neq\boldsymbol{\sigma}} Q(\boldsymbol{\sigma}''|\boldsymbol{\sigma})A(\boldsymbol{\sigma}''|\boldsymbol{\sigma}) & \text{if } \boldsymbol{\sigma}' = \boldsymbol{\sigma},
\end{cases}
\end{align}
where the acceptance probability is defined as
\begin{align}
\label{eq:MH_acceptance}
A(\boldsymbol{\sigma}'|\boldsymbol{\sigma}) = \min \left( 1, \frac{\pi(\boldsymbol{\sigma}')}{\pi(\boldsymbol{\sigma})} \frac{Q(\boldsymbol{\sigma}|\boldsymbol{\sigma}')}{Q(\boldsymbol{\sigma}'|\boldsymbol{\sigma})} \right).
\end{align}

The hybrid quantum-classical MCMC algorithm uses quantum dynamics to propose new transitions. 
Short-time evolution under Hamiltonians related to QA or QAOA for Ising models drives transitions toward the ground state and low-energy excited states, enabling efficient sampling from low-temperature Boltzmann distributions.
In the original work \cite{layden2023quantum} and its improved works \cite{PhysRevResearch.6.033105, arai2025quantum}, a symmetric operator $U = U^{\top}$ is chosen to simplify the evaluation of the acceptance probability, allowing transitions induced by quantum dynamics to be incorporated directly as steps of the MCMC process.
Notably, the original scheme introduced by Ref.~\cite{layden2023quantum} is termed the quantum-enhanced MCMC (Qe-MCMC).

An alternative approach employs a neural network as a surrogate model for the quantum dynamics in the proposal distribution \cite{nakano2025neuralnetworkassistedmontecarlosampling}. 
In this scheme, a classical neural network is trained to learn the output distribution of the quantum circuit, and the trained model is then used to generate proposals while the MCMC update is performed entirely classically. 
Because the proposal distribution is explicitly computable in this case, asymmetric QAOA circuits can be used as the underlying quantum generator for the proposal distribution.
A distribution estimator based on autoregressive neural networks, such as Masked Autoencoder for Distribution Estimation (MADE) \cite{germain2015made, uria2016neural}, is used to learn the output distribution \cite{10.21468/SciPostPhys.15.1.018}. 
In the remainder of this paper, this approach utilizing QAOA and MADE is referred to as the QAOA-assisted Neural Monte Carlo Sampler (QAOA-NMC).


\section{Degenerate ground-state sampling for small systems}
\label{sec:fair_sampling}
\begin{figure*}[htbp]
    \centering
    \includegraphics[width=\linewidth]{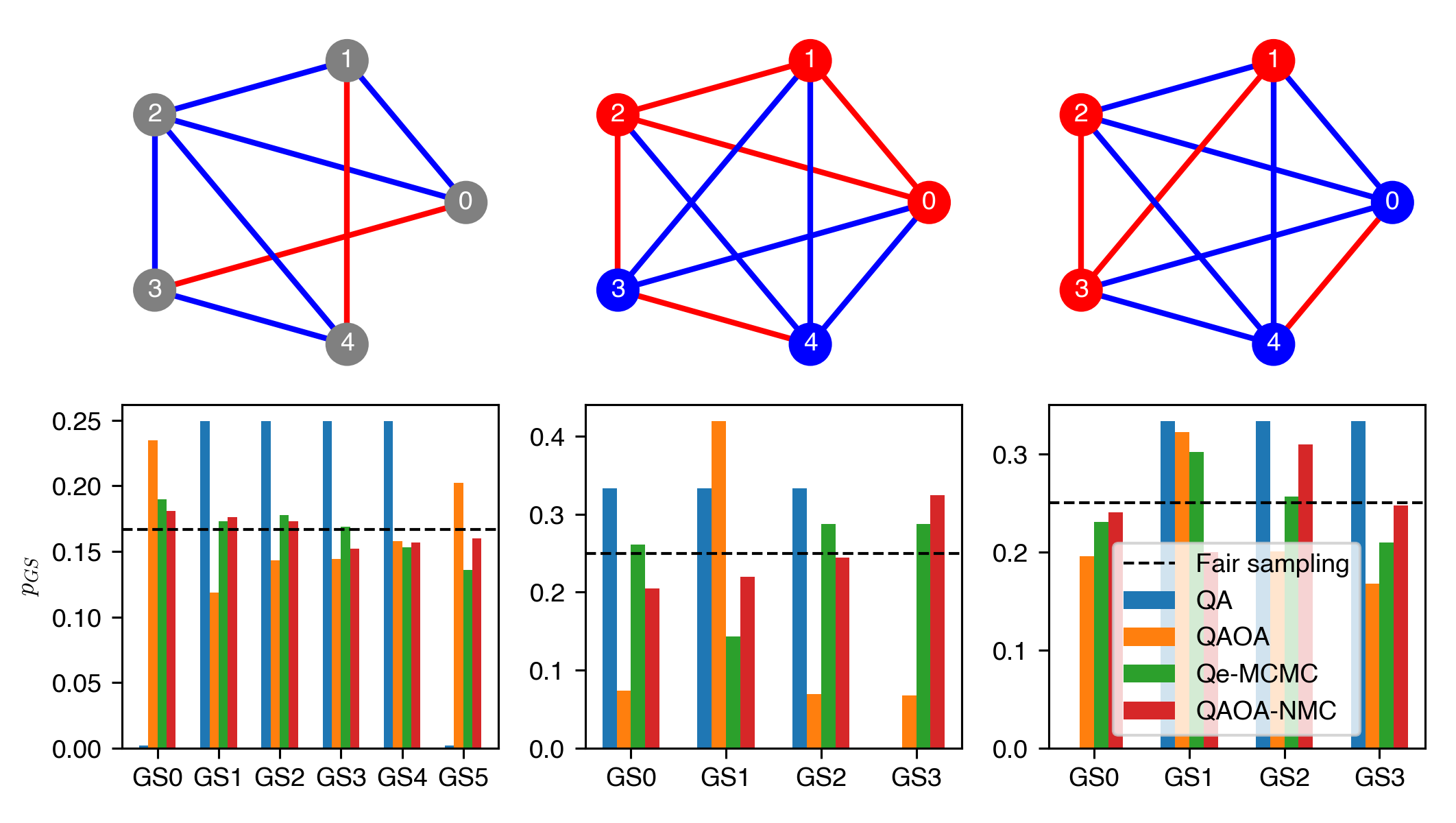}
    \caption{Three five-site Ising models with degenerate ground states and their corresponding sampling probabilities. \textbf{(Top)} Model specifications. Node colors represent the local fields $h_i$, and edge colors represent the couplings $ J_{ij}$: red indicates coefficients $+1$, blue indicates $-1$, and gray indicates $0$. \textbf{(Bottom)} Sampling probabilities over the degenerate ground states. The black dashed line indicates perfectly fair, that is, uniform, sampling over the ground-state manifold.}
    \label{fig:small_instance_histogram}
\end{figure*}
\begin{figure}[htbp]
    \centering
    \includegraphics[width=\linewidth]{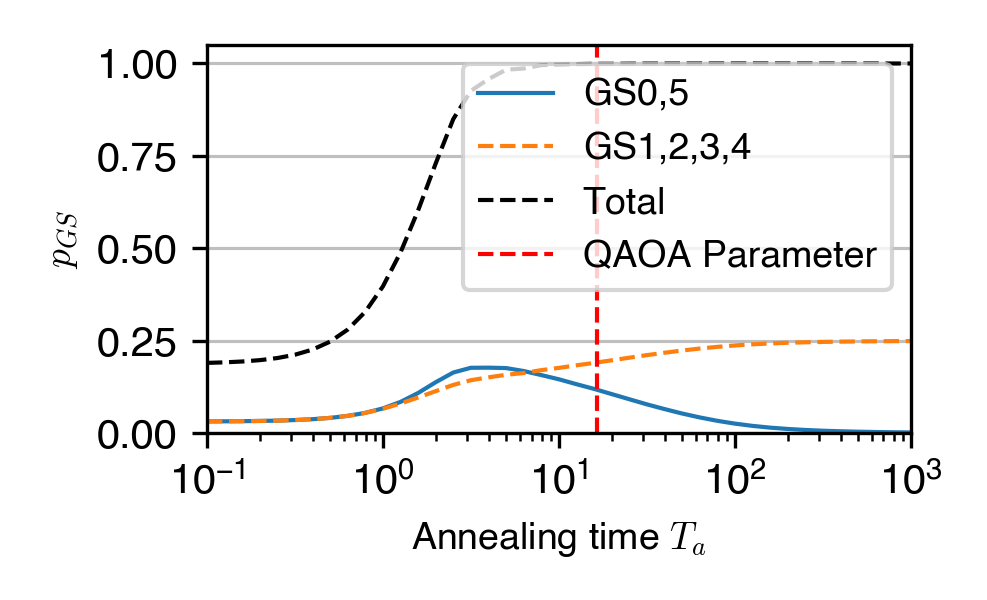}
    \caption{Dependence of the final QA state on the annealing time $T_a$ for the instance with sixfold degeneracy shown in Fig.~\ref{fig:small_instance_histogram}. The red line indicates the total effective evolution time, $T_{\mathrm{QAOA}} = \sum_{i=1}^{p} \left( \beta_i + \gamma_i \right)$ obtained by interpreting the optimized QAOA parameters as effective evolution times.}
    \label{fig:small_instance_evolution_time}
\end{figure}
To investigate the sampling properties of quantum heuristics for the Ising model, we first perform numerical simulations on small-scale models with degenerate ground states. The models considered are $k = 2$ Ising models with five sites. The interaction coefficients take only two values, $\{+1, -1 \}$, which results in a degenerate ground-state manifold.

We sample degenerate ground states using four methods: direct sampling of the final state via Quantum Annealing (QA) and QAOA, and two hybrid quantum-classical MCMC approaches, Qe-MCMC and QAOA-NMC. 
For all methods, a transverse-field driver Hamiltonian is employed as the mixer Hamiltonian. 
The specific settings are as follows: For QA, we calculate the final state by integrating the Schrödinger equation using the QuTiP library with an annealing time $T_a=10^3$. 
For QAOA, we employ a circuit of depth $p=5$, optimize parameters via the BFGS algorithm \cite{byrd1995limited}, and perform $10^3$ measurements. 
QAOA-NMC uses these samples as training data, with all hyperparameters and settings adopted from Ref.~\cite{nakano2025neuralnetworkassistedmontecarlosampling}. 
Qe-MCMC is executed with hyperparameters based on Ref.~\cite{christmann2024quantum}. 
For both MCMC approaches, $10^3$ samples are drawn from the Boltzmann distribution with $\beta=10$.
Regarding implementation, Qulacs \cite{suzuki2021qulacs} is used for quantum circuit simulations of QAOA and Qe-MCMC, while PyTorch \cite{NEURIPS2019_bdbca288} is used for the training and inference of MADE.

Figure~\ref{fig:small_instance_histogram} shows the sampling results for the three models previously examined in Ref.~\cite{PhysRevA.100.030303, PhysRevE.111.054103}. 
As confirmed in earlier studies, QA suppresses the occurrence probabilities of certain ground states, even for sufficiently large $T_a$. 
Interestingly,
in the case of short-time annealing ($T_a \lesssim 10$), the sampling probabilities of all ground states remain fair in the final state (Fig.~\ref{fig:small_instance_evolution_time}).
This suggests that, in principle, fair sampling might be achieved by appropriately optimizing the annealing time. However, determining such an optimal value of $T_a$ in advance is highly nontrivial.

By contrast, hybrid quantum-classical MCMC methods show no such suppression, retaining sampling probabilities for all degenerate ground states. 
This is primarily because both algorithms employ QA Hamiltonian dynamics with short evolution times for transitions, thereby mitigating the driver-induced suppression of degenerate ground states. 
In fact, the red dotted line in Fig.~\ref{fig:small_instance_evolution_time} denotes $T_{\rm{QAOA}} = \sum_{i=1}^{p} \left( \beta_i + \gamma_i \right)$, the effective annealing time derived from optimized QAOA parameters, which corresponds to the temporal region where annealing maintains relatively fair sampling.
Furthermore, the Metropolis--Hastings post-processing applied to the quantum proposals also contributes significantly to restoring fair sampling. 
QAOA-NMC, in particular, shows reduced bias compared with the raw QAOA output distribution on which it is trained. This improvement arises because the low-temperature Boltzmann distribution imposes acceptance probabilities that effectively act as a filter, allowing only ground states and a subset of low-energy excited states to be accepted. These findings indicate that combining quantum heuristics with MCMC is a promising strategy for achieving fair sampling.


\section{Random $k$-SAT sampling}
\label{sec:random_sat_sampling}
In this section, we perform a numerical analysis of the fair sampling and applications of hybrid quantum-classical MCMC on random k-SAT instances.

\subsection{Random $k$-SAT}
The Boolean Satisfiability Problem (SAT) is one of the most fundamental problems in computer science. 
It asks whether there exists an assignment of variables that evaluates a given logical formula to \texttt{True}. 
In particular, the case where each constraint consists of a logical disjunction (\texttt{OR}) of $k$ literals is referred to as $k$-SAT.

For $N$ binary variables $x_1, \dots, x_N$, let a literal $l$ be either the variable itself $x_i$ or its negation $\neg x_i$. 
An instance of $k$-SAT is described in the conjunctive normal form (CNF) as a logical conjunction (\texttt{AND}) of "clauses" $C_m$, each consisting of the disjunction of $k$ literals:
\begin{align}
F(\boldsymbol{x}) = C_1 \land C_2 \land \dots \land C_M = \bigwedge_{m=1}^{M} \left( \bigvee_{j=1}^{k} l_{m,j} \right).
\end{align}
Here, $M$ denotes the total number of clauses, and $l_{m,j}$ represents the $j$-th literal included in the $m$-th clause. 
The objective of the problem is to find a configuration of variables $\boldsymbol{x} \in \{0, 1\}^N$ such that $F(\boldsymbol{x}) = 1$ (\texttt{True}).

The $k$-SAT problem addressed in this study is a typical class of Constraint Satisfaction Problems (CSP). 
From a physical perspective, the SAT problem can be formulated as an energy minimization problem by treating each clause $C_m$ as a local interaction energy term. 
We define a Hamiltonian $H_m$ for each clause, which assigns an energy of $0$ if the logical formula is satisfied \texttt{True} and a positive energy penalty $H_m > 0$ if it is not satisfied \texttt{False}. 
The total Hamiltonian of the system, $H_{\rm{SAT}}$, is then constructed as:
\begin{align}
H_{\rm{SAT}} = \sum_{m=1}^{M} H_m(\boldsymbol{x})
\end{align}
In $k$-SAT, the sole condition under which a clause $C_m$ remains unsatisfied is when all $k$ literals in the clause evaluate to \texttt{False}. Therefore, by constructing a penalty function that increases the energy only under this condition and expanding it using Ising variables $s_i \in \{+1, -1\}$ (or $x_i \in \{0, 1\}$), the problem can be mapped to the Ising model or QUBO formulation introduced in the previous section.

For example, the Hamiltonian for a clause $C = x_i \lor x_j \lor x_k$ in 3-SAT (where variables are binary) can be constructed as follows. 
Since this clause evaluates to \texttt{False} only when $(x_i, x_j, x_k) = (0, 0, 0)$, the penalty term is given by:
\begin{align}
H_C = (1-x_i)(1-x_j)(1-x_k).
\end{align}
This term evaluates to $1$ only for the state $(0,0,0)$ and $0$ otherwise. 
By expanding this expression and converting the variables to spin variables via $s_i = 1-2x_i$, an Ising Hamiltonian containing up to three-body interactions is obtained.

Furthermore, it is known that the computational hardness of $k$-SAT changes dramatically depending on the constraint density $\alpha = M/N$, defined as the ratio of the number of clauses $M$ to the number of variables $N$. 
A phase transition regarding the existence of a solution (SAT/UNSAT transition) occurs near a specific critical value $\alpha_c$. 
This region is of particular importance as it represents the most difficult regime even for classical heuristics.

\subsection{Method}
\subsubsection{Instance Set}
\begin{figure}
    \centering
    \includegraphics[width=\linewidth]{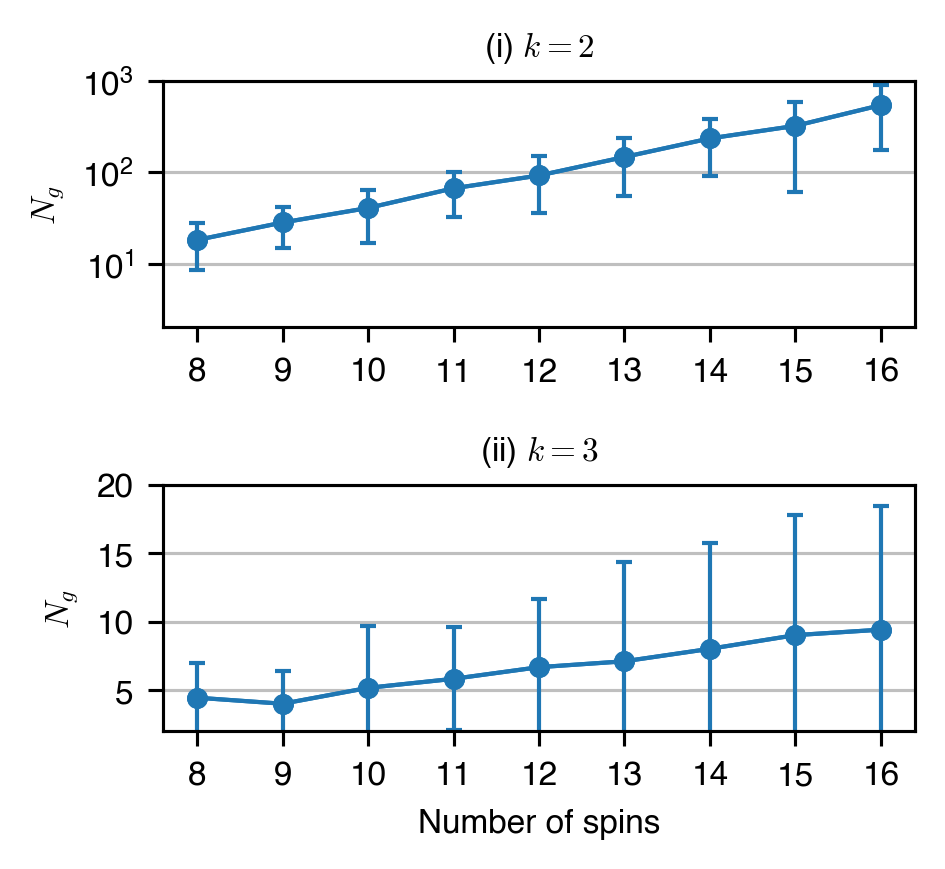}
    \caption{Number of solutions, or degeneracy, $N_g$, of the generated random 2-SAT and 3-SAT instances as a function of the problem size $N$.}
    \label{fig:instance_set}
\end{figure}
In this study, we utilize Random $k$-SAT problems with $k=2$ and $3$. 
Specifically, to address problems with a constraint density near the critical point $\alpha_c$ \cite{PRXQuantum.5.030348}, where quantum heuristics are expected to be particularly effective, and to ensure the existence of multiple SAT solutions, we generate the instance set as follows. 
First, we generate a Random $k$-SAT instance consisting of $M = \lfloor \alpha_c N \rfloor + 1$ clauses, randomly selected from the set of all possible ${N \choose k} 2^k$ clauses. Next, we use a classical solver to filter out instances that have one or zero SAT solution.
Through this procedure, we prepare an instance set consisting of 100 random instances for each problem size $N$, ranging from $8 \leq N \leq 16$. 
Figure~\ref{fig:instance_set} shows the number of degenerate ground states, $N_g$, for each problem size $N$.

\subsubsection{Sampling}
In this experiment, we perform the sampling of SAT solutions for the generated instance set using a hybrid quantum-classical MCMC. 
We adopt QAOA-NMC for this purpose. 
The rationale for this choice is twofold: QAOA-NMC using an optimized QAOA circuit is expected to demonstrate faster convergence than Qe-MCMC, and it facilitates a direct comparison with the QAOA output distribution. 
This approach allows us to clearly elucidate the correction effects resulting from the MCMC steps.

The detailed settings for QAOA-NMC are described below. 
For QAOA, we use a circuit with depth $p=5$.
To reduce the computational cost associated with optimization, we introduce linear parameters \cite{sakai2024linearlysimplifiedqaoaparameters}. 
This involves replacing the $2p$ parameters of the QAOA with a linear schedule consisting of four parameters: $\beta_l = \beta_{\rm{slope}} \frac{l}{p} + \beta_{\rm{intcp}}$ and $\gamma_l = \gamma_{\rm{slope}} \frac{l}{p} + \gamma_{\rm{intcp}}$. 
Furthermore, since such linear schedules contain parameters (referred to as ``fixed angles'') that universally yield good cost functions for instances, we prepare two types of QAOA circuits: one based on optimized parameters and another based on these fixed parameters derived from the schedule \cite{ichikawa2025optimalelementalconfigurationsearch, Montanez-Barrera2025}.
For MADE, the size of the training dataset is set to $10^3$, and all hyperparameters are identical to those in Ref.~\cite{nakano2025neuralnetworkassistedmontecarlosampling}.
Sampling of the ground state is performed via the Boltzmann distribution with $\beta=10$. 
We executed $10^4$ sampling steps for each of 10 random initial states.

\subsubsection{Hybrid Algorithm}
As shown in Fig.~\ref{fig:instance_set}, the instance set exhibits a high degree of degeneracy. Particularly for $k=2$, the system possesses a vast number of solutions; consequently, it is highly probable that some ground states are missing from the training data, and the learned MADE may fail to transition to all ground states. 
To correct these biases, we introduce a new hybrid method that combines the MADE update with a sweep consisting of $N$ single spin-flip (SSF) updates.
The computational cost of an SSF sweep is minimal, and the hybrid update remains of the same order as the MADE update. 
The introduction of the SSF sweep is intended to explore other ground states located in the vicinity of the state proposed by MADE.
Since many instances in $k=2$ form clusters with close Hamming distances, this approach is expected to play a role in complementing the sampling of ground states that were missing during the MADE training.
Hereafter, we denote this hybrid variant combining QAOA-NMC and SSF sweeps as QAOA-assisted Hybrid neural MCMC (QAOA-HMC).

\subsubsection{Classical Algorithms}
As benchmarks for classical algorithms in this study, we adopt PT-ICM, a state-of-the-art sampling algorithm for 2-body Ising models, and WalkSAT, a representative method for Stochastic Local Search (SLS).

\paragraph{PT-ICM}
PT-ICM \cite{zhu2015efficient} is an algorithm that combines Parallel Tempering (PT) \cite{PhysRevLett.57.2607, doi:10.1143/JPSJ.65.1604} with Isoenergetic Cluster Moves (ICM) \cite{houdayer2001cluster}. 
PT simulates multiple replicas with different temperature parameters in parallel and periodically exchanges their configurations. 
High-temperature replicas perform global exploration, while low-temperature replicas perform local optimization, enabling the system to efficiently overcome energy barriers. 
The ICM identifies clusters of interacting spins and performs non-local updates by flipping them collectively. 
This update is designed to preserve the total energy of the system (or cause minimal change), facilitating transitions to distant valleys in the state space by ``tunneling'' through regions with high energy barriers. 
The overall process is described by the procedure below. 
Due to the synergistic effect of PT and ICM, PT-ICM achieves extremely fast relaxation and efficient sampling even in complex systems like spin glasses.

\paragraph{WalkSAT}
WalkSAT \cite{10.5555/1597148.1597256} is an incomplete solver that heuristically explores the solution space, unlike complete solvers such as the DPLL (Davis-Putnam-Logemann-Loveland) algorithm \cite{10.1145/368273.368557}, which performs an exhaustive search. 
It is known for its ability to discover satisfying assignments extremely quickly, especially for random $k$-SAT instances.

The search process of WalkSAT begins with a random initial assignment and proceeds by repeating the following steps:
\begin{enumerate}
\item Randomly select one unsatisfied clause under the current variable assignment.
\item Determine which variable within that clause to flip. At this point, one of the following operations is performed based on a probability $p$ (noise parameter):
\begin{itemize}
    \item Random Move (probability $p$): Randomly select a variable in the clause and flip its value. This promotes escape from local optima (local minima).
    \item Greedy Move (probability $1-p$): Select the variable that, when flipped, results in the greatest reduction (or minimal increase) in the total number of unsatisfied clauses.
\end{itemize}
\end{enumerate}

By balancing this "greedy optimization" and "random walk," WalkSAT efficiently explores the complex energy landscape, aiming to reach a global optimum (satisfying assignment). 
In this study, we utilize an algorithm called WalkSATlm \cite{10.1093/comjnl/bxu135}, which is modified with a more complex cost function. 
Although WalkSAT is originally designed to find a single SAT solution, it can be utilized to enumerate SAT solutions with a simple modification. 
This involves adding a new clause to the problem's CNF that prohibits the already discovered solution, thereby searching for undiscovered solutions. 
By repeating this process until the CNF becomes UNSAT, all SAT solutions can be enumerated.
\begin{figure*}[htbp]
    \centering
    \includegraphics[width=\linewidth]{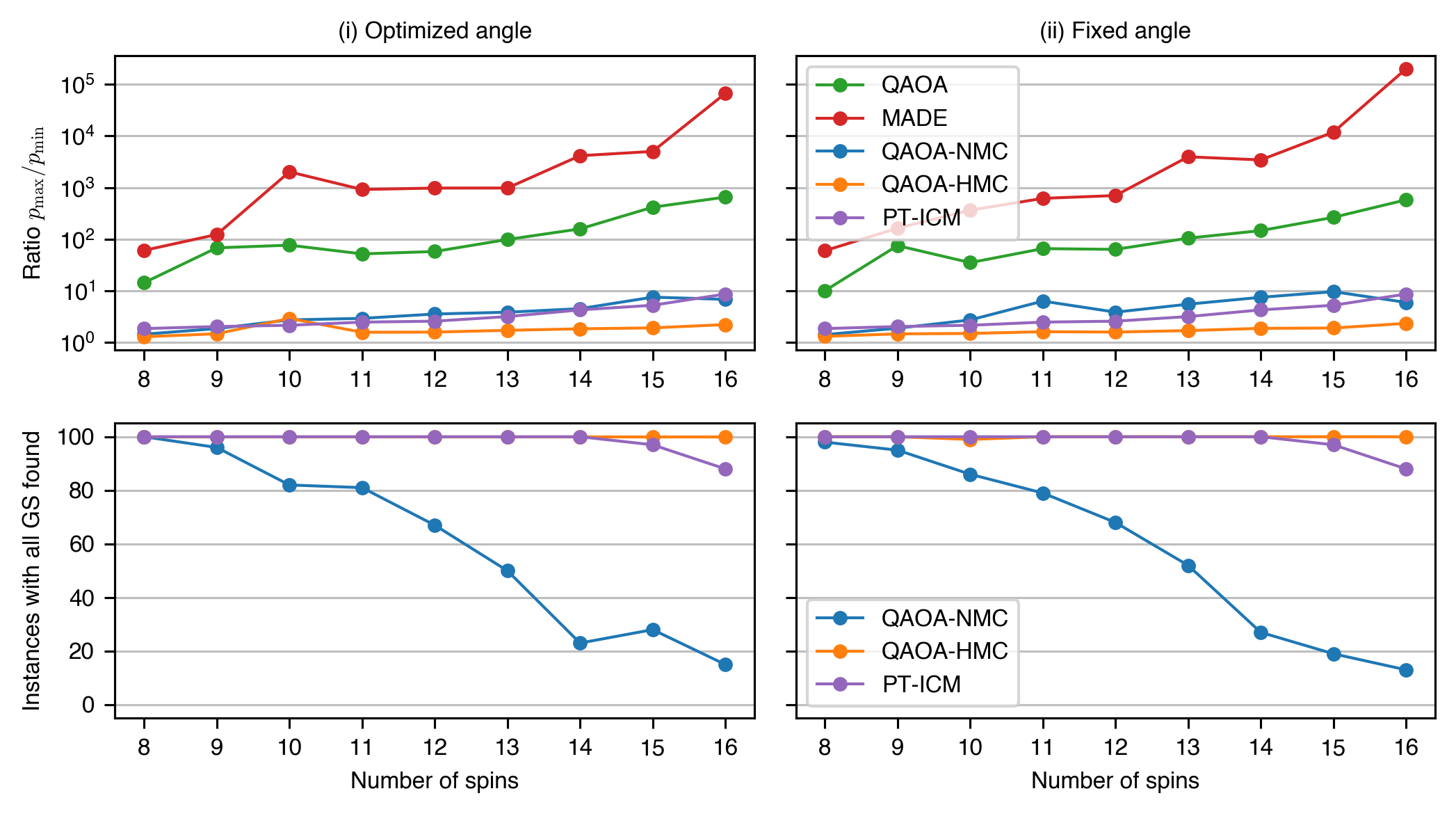}
    \caption{The ratio $p_{\max}/p_{\min}$ and the number of instances where all ground states were successfully discovered as a function of the problem size $N$ for the $k=2$ case.}
    \label{fig:k2_fair_sampling}
\end{figure*}

\subsection{Fair Sampling}
\begin{figure*}[htbp]
    \centering
    \includegraphics[width=\linewidth]{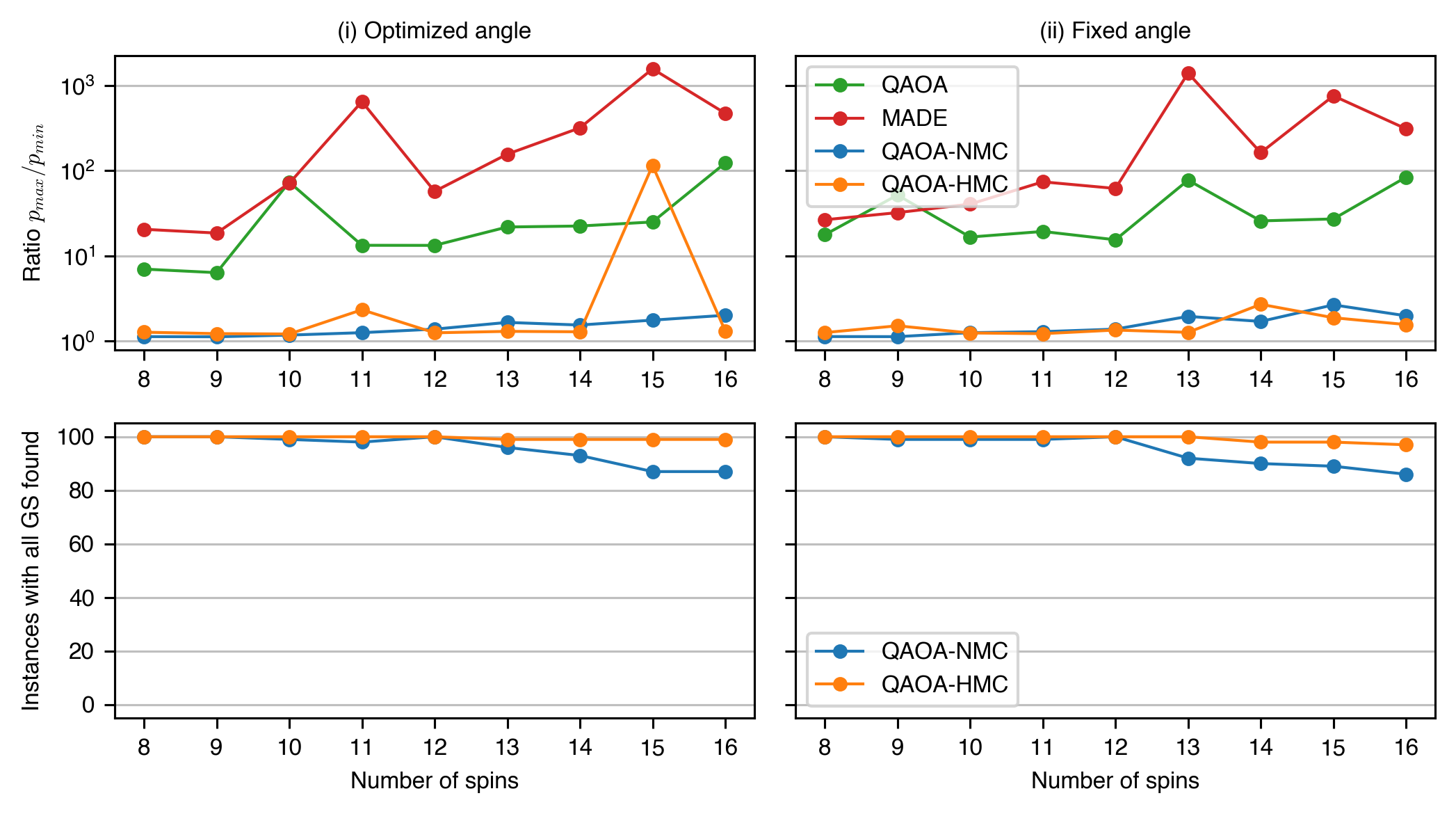}
    \caption{The ratio $p_{\max}/p_{\min}$ and the number of instances where all ground states were successfully discovered as a function of the problem size $N$ for the $k=3$ case.}
    \label{fig:k3_fair_sampling}
\end{figure*}
First, to assess the fair sampling capability, we quantify the sampling bias using the ratio of the maximum to minimum probabilities of the ground states, denoted as $p_{\max}/p_{\min}$.
For QAOA and MADE, this ratio is derived based on the directly computed output distributions; for MCMC, it is calculated from the histogram constructed from all sampled ground states. 
It should be noted that for MCMC, there are cases where the sampler fails to visit all degenerate ground states. 
In such instances, the ratio $p_{\max}/p_{\min}$ cannot be defined, and thus these results are excluded from the analysis.

\subsubsection{Random 2-SAT}
Figure~\ref{fig:k2_fair_sampling} illustrates the ratio $p_{\max}/p_{\min}$ and the number of instances where all ground states were successfully discovered as a function of the problem size $N$ for the $k=2$ case. 
As discussed in the analysis of small systems, optimized QAOA exhibits significant bias. 
Consequently, the MADE trained on these outputs becomes an even more biased sampler. 
Therefore, QAOA-NMC often fails to sample all degenerate ground states. 
In contrast, QAOA-HMC successfully discovered all ground states for all instances. 
Furthermore, in terms of the ratio $p_{\max}/p_{\min}$, it demonstrates fairness comparable to, or even surpassing, that of PT-ICM, which is known for achieving fair sampling. 
Additionally, QAOA yielded nearly identical results regardless of whether optimized parameters or fixed angles were used.
This is because the linear schedule employed in this study exhibits a strong concentration of optimal parameters \cite{sakai2024linearlysimplifiedqaoaparameters}, and the fixed angles, derived as the medians of these values, are very close to the optimal parameters. 
While this does not directly affect fair sampling performance, the use of fixed angles eliminates the need for the computationally expensive optimization of QAOA, making it a practically significant property.

\subsubsection{Random 3-SAT}
Figure~\ref{fig:k3_fair_sampling} presents the ratio $p_{\max}/p_{\min}$ and the number of instances where all ground states were discovered as a function of problem size $N$ for $k=3$. 
While the trend is fundamentally similar to that of $k=2$, the overall degree of bias is reduced. 
This is attributed to the smaller total number of solutions, $N_g$, compared to the $k=2$ case. 
Consequently, even QAOA-NMC succeeded in exploring all ground states for many instances. 
However, in the case of $k=3$, there were instances where even QAOA-HMC failed. 
All such instances possess configurations that are weakly correlated with other ground states; thus, they cannot benefit from the local neighborhood search provided by the SSF sweep.

\subsection{Model Counting}
\begin{figure*}[htbp]
    \centering
    \includegraphics[width=\linewidth]{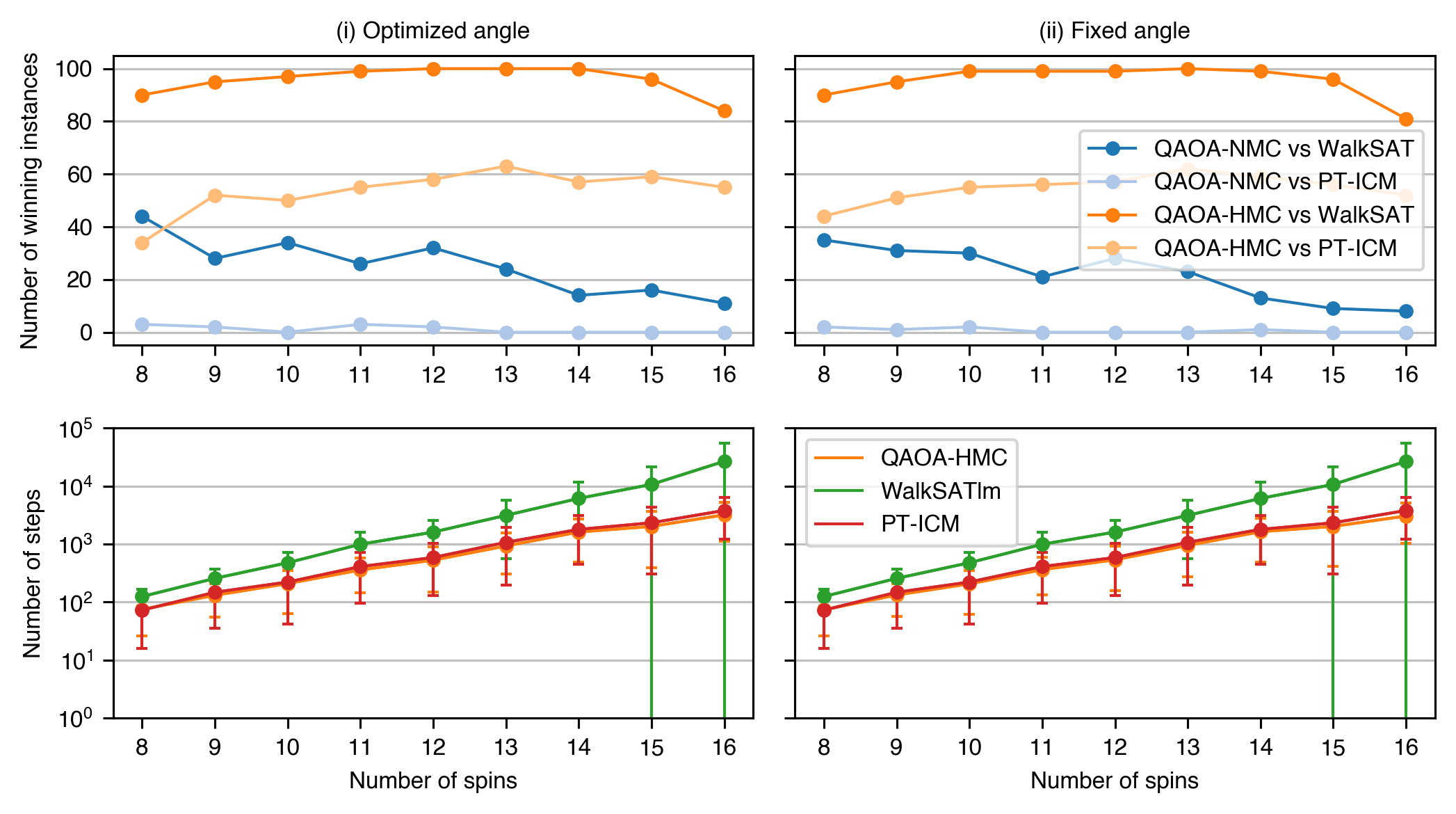}
    \caption{The number of steps required to enumerate all ground states and the number of instances where MCMC demonstrated superiority over WalkSATlm and PT-ICM, as a function of problem size $N$ for $k=2$.}
    \label{fig:k2_model_counting}
\end{figure*}
\begin{figure*}[htbp]
    \centering
    \includegraphics[width=\linewidth]{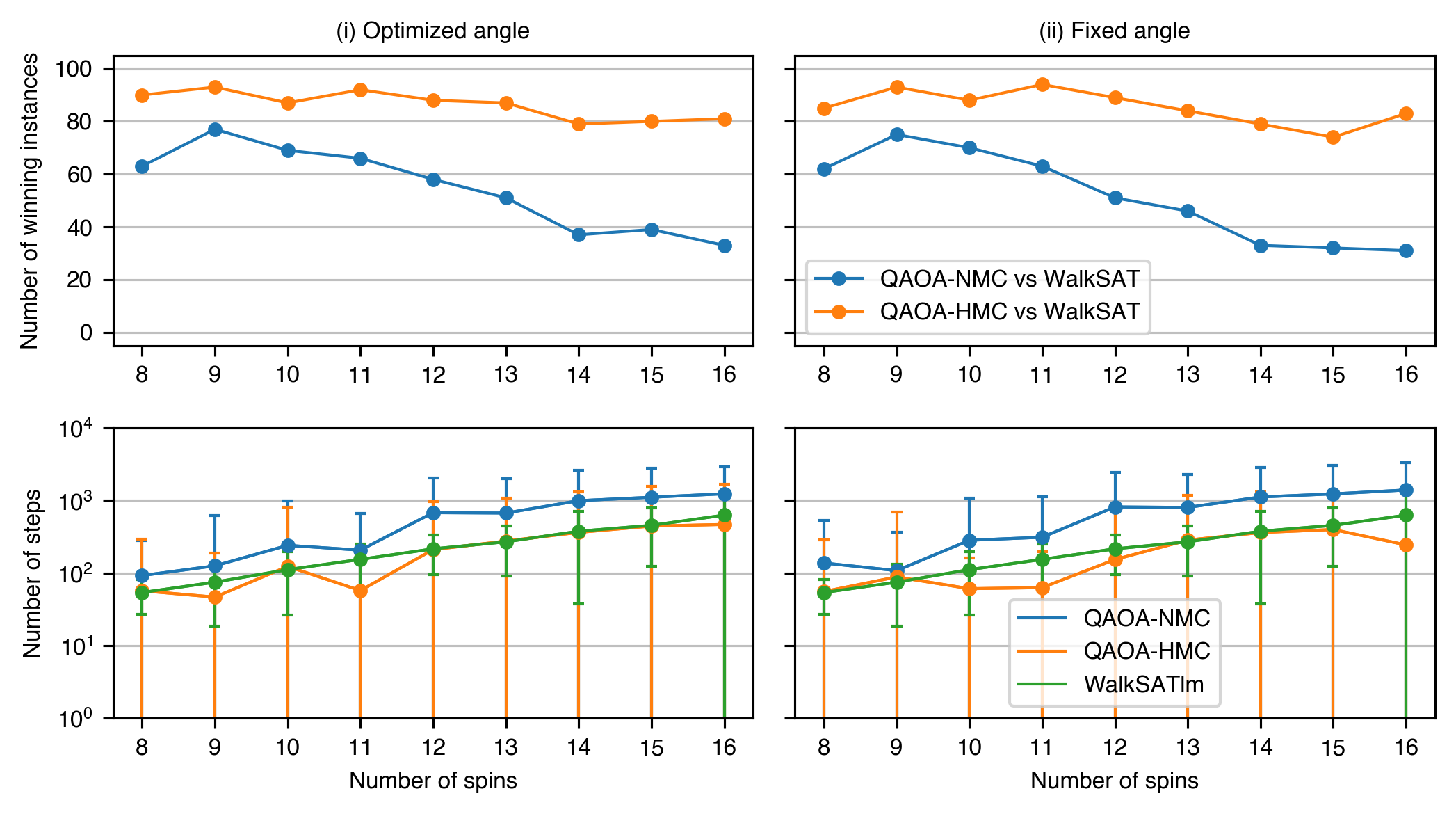}
    \caption{The number of steps required to enumerate all ground states and the number of instances where MCMC demonstrated superiority over WalkSATlm, as a function of problem size $N$ for $k=3$.}
    \label{fig:k3_model_counting}
\end{figure*}
Next, as an application of the fair sampling property, we perform the enumeration of SAT solutions. 
This is a problem known as Model Counting (\#SAT) \cite{VALIANT1979189}. 
Here, we compare the counting performance between MCMC sampling and WalkSATlm. 
We define a state transition as a single step and evaluate the total number of steps required to discover all solutions. 
For QAOA-NMC, QAOA-HMC and WalkSATlm, we calculate the average number of steps over 10 trials, whereas for PT-ICM, we count the steps in a single trial. 
As in the previous section, instances where not all degenerate ground states were discovered within a limited number of steps are excluded from the analysis.

\subsubsection{Random 2-SAT}
Figure~\ref{fig:k2_model_counting} displays the number of steps required to enumerate all ground states and the number of instances where MCMC demonstrated superiority over WalkSATlm, as a function of problem size $N$ for $k=2$. 
QAOA-NMC fails to surpass the performance of PT-ICM in almost all instances and is also inferior to WalkSATlm in many cases. 
This is because, as mentioned in the previous section, QAOA-NMC fails to sample all ground states for many instances. 
Consequently, the step count data for QAOA-NMC is omitted from Fig.~\ref{fig:k2_model_counting}. 
In contrast, QAOA-HMC outperforms WalkSATlm in most cases and achieves performance nearly equivalent to that of PT-ICM.

\subsubsection{Random 3-SAT}
Figure~\ref{fig:k3_model_counting} displays the number of steps required to enumerate all ground states and the number of instances where MCMC was superior, as a function of problem size $N$ for $k=3$. 
Unlike the $k=2$ case, QAOA-NMC possesses a certain degree of competitiveness against WalkSATlm. 
Although QAOA-HMC still outperforms WalkSATlm in many instances, the scaling of the average step count is comparable to that of WalkSATlm. 
As discussed in the previous section, the efficacy of the SSF sweep diminishes in this regime; consequently, the bias of QAOA-HMC increases, leading to significant performance variance across instances.


\section{conclusion}
\label{sec:conclusion}
In this work, we investigated the fair sampling property of hybrid quantum-classical MCMC algorithms for problems with degenerate ground states. Our focus is not on demonstrating a quantum advantage, but rather on clarifying whether biased sampling induced by quantum heuristics can be mitigated within a principled sampling framework.

Through numerical experiments on small Ising models, we showed that although quantum annealing and QAOA with transverse-field drivers exhibit strong sampling bias, the introduction of Metropolis--Hastings post-processing substantially restores uniformity over degenerate ground states. This result highlights that sampling bias arising from quantum dynamics does not necessarily preclude fair sampling when combined with classical MCMC updates that enforce detailed balance.

We further examined random $k$-SAT problems as practically relevant benchmarks with highly degenerate solution spaces. For random 2-SAT, we found that a hybrid MCMC combining QAOA-trained neural proposals with single spin-flip updates achieves a level of fairness comparable to that of PT-ICM, a well-established classical fair sampler. For random 3-SAT, where PT-ICM is no longer applicable, the hybrid approach still yields approximately uniform sampling, although instance-dependent limitations remain, particularly for solutions that are weakly connected in configuration space.

As an application, we studied model counting and observed that the number of transitions required to enumerate all solutions using hybrid MCMC is comparable to that of WalkSAT, a representative stochastic local search algorithm. This suggests that hybrid MCMC can serve as an alternative sampling-based approach for enumeration tasks, rather than a replacement for existing classical solvers.

\begin{acknowledgments}
Y.N. would like to thank Ken N. Okada for valuable and insightful discussions.
This work is supported by MEXT Quantum Leap Flagship Program (MEXT Q-LEAP) Grant No. JPMXS0120319794, JST COI-NEXT Grant No. JPMJPF2014, and JST CREST JPMJCR24I3.
Y.N. is also supported by Grant-in-Aid for JSPS Fellows Grant No. 24KJ1606.
\end{acknowledgments}

\bibliographystyle{apsrev4-2}
\bibliography{cite}

\end{document}